\documentclass[aps,prd,amsmath,amssymb,superscriptaddress,preprintnumbers,twocolumn,10pt,nofootinbib,showkeys]{revtex4-1}
\usepackage{graphicx}
\usepackage{dcolumn}
\usepackage{bm}
\usepackage{amssymb}
\usepackage{latexsym}
\usepackage{booktabs}
\usepackage{amsmath}
\usepackage{multirow}
\usepackage{url}
\usepackage{footnote}
\usepackage{float}
\usepackage{threeparttable}
\usepackage[bottom]{footmisc}

\usepackage[normalem]{ulem}
\usepackage{color}
\usepackage{array}
\usepackage{enumerate}
\usepackage{adjustbox}
\usepackage{makecell}
\usepackage{diagbox}
\usepackage{epstopdf}
\usepackage{epsfig}
\usepackage{longtable}
\usepackage{supertabular}
\usepackage{algorithm}
\usepackage{pifont}
\usepackage{algorithmic}
\usepackage{changepage}
\usepackage{setspace}
\usepackage[colorlinks=true, linkcolor=blue, citecolor=blue]{hyperref}

\begin{document}

\title{Cosmological preference for a positive neutrino mass at 2.7$\sigma$: A joint analysis of DESI DR2, DESY5, and DESY1 data}

\author{Guo-Hong Du}
\affiliation{Liaoning Key Laboratory of Cosmology and Astrophysics, College of Sciences, Northeastern University, Shenyang 110819, China}

\author{Tian-Nuo Li}
\affiliation{Liaoning Key Laboratory of Cosmology and Astrophysics, College of Sciences, Northeastern University, Shenyang 110819, China}

\author{Peng-Ju Wu}
\affiliation{School of Physics, Ningxia University, Yinchuan 750021, China}

\author{Jing-Fei Zhang}
\affiliation{Liaoning Key Laboratory of Cosmology and Astrophysics, College of Sciences, Northeastern University, Shenyang 110819, China}

\author{Xin Zhang}\thanks{Corresponding author}\email{zhangxin@mail.neu.edu.cn}
\affiliation{Liaoning Key Laboratory of Cosmology and Astrophysics, College of Sciences, Northeastern University, Shenyang 110819, China}
\affiliation{MOE Key Laboratory of Data Analytics and Optimization for Smart Industry, Northeastern University, Shenyang 110819, China}
\affiliation{National Frontiers Science Center for Industrial Intelligence and Systems Optimization, Northeastern University, Shenyang 110819, China}

\begin{abstract}

Neutrinos and dark energy (DE) have entered a new era of investigation, as the latest DESI baryon acoustic oscillation measurements tighten the constraints on the neutrino mass and suggest that DE may be dynamical rather than a cosmological constant. In this work, we obtain a high-confidence measurement of the neutrino mass within a dynamical DE framework. A joint analysis of DESI DR2, cosmic microwave background, DESY5 supernova, and DESY1 weak lensing data yields a total neutrino mass of $\sum m_\nu = 0.098^{+0.016}_{-0.037}\,\mathrm{eV}$, indicating a measurement for a non-zero, positive neutrino mass at the $2.7\sigma$ level within the $w_0w_a$CDM framework. This high-confidence measurement is driven mainly by these factors: (i) the DESI's preference for a dynamical DE with its equation of state evolving from $w< -1$ at early times to $w> -1$ at late times, thus leading to a larger neutrino mass; (ii) treating $N_{\mathrm{eff}}$ as a free parameter together with the inclusion of weak lensing data, which likewise allows for an increased neutrino mass. In the future, even higher-confidence measurements of neutrino mass are expected with stronger preferences for dynamical DE in light of more complete DESI data releases.

\end{abstract}
\maketitle

\section{Introduction}

Neutrino oscillation experiments have provided compelling evidence that neutrinos have non-zero masses, constituting the only experimentally established manifestation to date of physics beyond the Standard Model. Consequently, measuring the absolute neutrino mass has become a central objective in both particle physics and cosmology. On one hand, in particle physics, oscillation data have measured the two mass-squared differences among the three neutrino mass eigenstates $ m_1, m_2, m_3 $, with solar neutrino measurements yielding $ \Delta m_{21}^2 \approx 7.4 \times 10^{-5}~\text{eV}^2 $ and atmospheric neutrino experiments giving $ |\Delta m_{32}^2| \approx 2.4 \times 10^{-3}~\text{eV}^2 $ \cite{ParticleDataGroup:2024cfk}. However, oscillation experiments cannot determine the absolute neutrino mass, and the unknown sign of $ \Delta m_{32}^2 $ implies two possible neutrino mass hierarchies, namely normal hierarchy (NH) with $m_1< m_2\ll m_3$ and inverted hierarchy (IH) with $m_3\ll m_1<  m_2$. Oscillation data imply a lower bound on the total mass of $ \sum m_\nu \gtrsim 0.06\,{\rm eV} $ for NH and $ \sum m_\nu \gtrsim 0.10\,{\rm eV} $ for IH \cite{Esteban:2020cvm,deSalas:2020pgw}. Meanwhile, $\beta$-decay experiments provide direct laboratory measurements of the neutrino mass; the latest result from KATRIN gives $ \sum m_\nu < 1.35\,{\rm eV} $ \cite{KATRIN:2024cdt}. On the other hand, cosmological observations provide a highly sensitive probe of the neutrino masses. Neutrinos behave as relativistic radiation in the early universe and contribute to the matter density at late times, while their free-streaming motion suppresses the growth of small-scale structure. These effects imprint on the cosmic microwave background (CMB), big bang nucleosynthesis (BBN), and the large-scale structure of the universe \cite{Zhang:2014nta,Zhang:2014dxk,TopicalConvenersKNAbazajianJECarlstromATLee:2013bxd,Coulton:2018ebd,Zhang:2019ipd,Tanseri:2022zfe}. Analyses combining Planck 2018 CMB data and Sloan Digital Sky Survey baryon acoustic oscillations (BAO) data within the $\Lambda$-cold dark matter ($\Lambda$CDM) model yield an upper limit of $ \sum m_\nu < 0.12\,{\rm eV} $ at 95\% confidence level~\cite{Planck:2018vyg}, highlighting that cosmological observations currently provide far tighter constraints on neutrino mass than particle physics experiments.

However, cosmological measurements of neutrino mass depend sensitively on the cosmological model, especially on the nature of dark energy (DE) \cite{Zhang:2017rbg}. In our earlier systematic studies, we examined in detail the degeneracy between $\sum m_\nu$ and the DE equation-of-state (EoS) parameter $w$~\cite{Zhang:2015uhk,Zhao:2016ecj,Zhang:2017rbg}. We found that when the evolution of $w$ corresponds to quintessence, namely $w>-1$ at all times, or to a quintom scenario\footnote{The quintom denotes the DE behavior that $w$ crosses the cosmological constant boundary $w=-1$ during cosmic evolution \cite{Feng:2004ad}.} in which $w$ evolves from $w>-1$ at early times to $w<-1$ at late times, the upper limit on $\sum m_\nu$ becomes more stringent compared with $\Lambda$CDM. In contrast, when the evolution of $w$ corresponds to phantom, namely $w<-1$ at all times, or to a quintom scenario in which $w$ evolves from $w<-1$ at early times to $w>-1$ at late times, the upper limit on $\sum m_\nu$ is relaxed compared with $\Lambda$CDM. These results consistently indicate that once $w$ evolves from a larger value to a smaller value, the upper limit of $\sum m_\nu$ becomes tighter than in $\Lambda$CDM, while the opposite evolution leads to a looser limit\footnote{Similar conclusions were later reached by other works; see, e.g., Refs.~\cite{Yang:2017amu,RoyChoudhury:2018gay}.}.

Recently, the Dark Energy Spectroscopic Instrument (DESI) collaboration has published its second data release (DR2), incorporating BAO measurements from over 14 million extragalactic sources \cite{DESI:2025zgx}. The combination of DESI DR2 BAO with CMB and type Ia supernova (SN) data indicates a statistically significant deviation ($2.8 - 4.2\sigma$) from the $\Lambda$CDM paradigm, favoring a dynamically evolving DE model characterized by an EoS that evolves from $w < -1$ at early times to $w > -1$ at late times \cite{DESI:2025zgx}. These findings have stimulated numerous efforts to develop and test various explanations beyond $\Lambda$CDM, including various DE models \cite{Giare:2024smz,Li:2024qso,Sabogal:2024yha,Escamilla:2024ahl,Dinda:2024ktd,Li:2024qus,Huang:2025som,Wu:2025wyk,Li:2025ula,Li:2025ops,Barua:2025ypw,Yashiki:2025loj,Ling:2025lmw,Goswami:2025uih,Yang:2025boq,Pang:2025lvh,You:2025uon,Ozulker:2025ehg,Cheng:2025lod,Silva:2025twg,Chen:2025wwn,Zhang:2025dwu,Wang:2024dka,Li:2025owk,Pan:2025qwy,Lee:2025pzo,Li:2025vuh,Wu:2025vfs,Li:2026xaz,Li:2026hwq} and alternative cosmological scenarios \cite{Yang:2024kdo,RoyChoudhury:2024wri,Wu:2024faw,Du:2024pai,Jiang:2024viw,Ye:2024ywg,CosmoVerseNetwork:2025alb,Feng:2025mlo,Wang:2025dtk,Cai:2025mas,Li:2025cxn,Nojiri:2025low,Yang:2025ume,Kumar:2025etf,Li:2025eqh,RoyChoudhury:2025dhe,Li:2025htp,Giare:2025ath,Liu:2025myr,Du:2025csv,Yao:2025kuz,Cheng:2025cmb,Li:2025dwz,Bhattacharjee:2025xeb,Zhou:2025nkb,Braglia:2025gdo,Du:2025iow,Pan:2025psn,Odintsov:2025jfq,Song:2025bio,Afroz:2025iwo,Du:2026cly,FrancoAbellan:2026ori,Rodrigues:2026rgy,Ladeira:2026jne,Ivanov:2026dvl,Wang:2026sqy,Feng:2026pzs,Yang:2026yaq,Li:2026ldf}. Furthermore, DESI reports that in the $\Lambda$CDM model, the cosmological upper limit on the total neutrino mass is $\sum m_\nu < 0.064$~eV, which is close to the lower limits from oscillation experiments and in tension with the IH~\cite{DESI:2025zgx}. When considering dynamical DE models, the constraint relaxes to $\sum m_\nu < 0.16$~eV. This is due to the strong degeneracy between the EoS of DE and neutrino mass, and the DESI results favor the evolution of DE EoS crossing from $w<-1$ to $w>-1$, which leads to a larger neutrino mass, consistent with our earlier important studies~\cite{Zhang:2015uhk,Zhao:2016ecj,Zhang:2017rbg}. Meanwhile, it is worth noting that the DESI collaboration recently reported approximately a $2\sigma$ preference for a positive total neutrino mass, $\sum m_\nu = 0.106^{+0.050}_{-0.069}\,\mathrm{eV}$, within a DE framework mediated by the collapse of stars to cosmologically coupled black holes~\cite{DESI:2025ffm}. These results highlight the crucial role of DE dynamics in cosmological measurements of neutrino mass.

Overall, earlier studies~\cite{Zhang:2015uhk,Zhao:2016ecj,Zhang:2017rbg} showed that when DE evolves dynamically with the EoS transitioning from $w<-1$ at early times to $w>-1$ at late times, the neutrino mass limit is significantly relaxed. The current DESI data provide strong evidence for such dynamical behavior of DE, naturally favoring a larger neutrino mass. Based on this, we include both key neutrino parameters, $\sum m_\nu$ and $N_{\rm eff}$, in our analysis, and combine CMB, DESI BAO, DESY5 SN, and weak lensing data to achieve a high-confidence measurement of $\sum m_\nu$ at the $2.7\sigma$ level, accompanied by a detailed and systematic discussion of the impact of dynamical DE on neutrino mass measurement.

\section{Methodology and data}

\begin{table}[!htbp]
\caption{Flat priors on the cosmological parameters constrained in this work.}
\begin{center}
\renewcommand{\arraystretch}{1.3}
\begin{tabular}{@{\hspace{0.4cm}}c@{\hspace{0.5cm}} c@{\hspace{0.5cm}} c @{\hspace{0.4cm}} }
\hline\hline
\textbf{Model}       & \textbf{Parameter}       & \textbf{Prior}\\
\hline
$\Lambda$CDM        & $\Omega_{\rm b} h^2$                     & $\mathcal{U}$[0.005\,,\,0.1] \\
                    & $\Omega_{\rm c} h^2$                     & $\mathcal{U}$[0.01\,,\,0.99] \\
                    & $H_0$                                    & $\mathcal{U}$[20\,,\,100] \\
                    & $\tau$                                   & $\mathcal{U}$[0.01\,,\,0.8] \\
                    & $\ln(10^{10}A_{\rm s})$                 & $\mathcal{U}$[2.6\,,\,3.5] \\
                    & $n_{\rm s}$                              & $\mathcal{U}$[0.9\,,\,1.1] \\
\hline  
$w_0w_a$CDM                 & $w_0$                                    & $\mathcal{U}$[-3\,,\,1] \\
                    & $w_a$                                    & $\mathcal{U}$[-3\,,\,2] \\
\hline
Neutrino Parameter  & $\sum m_\nu$                             & $\mathcal{U}$[0\,,\,5] \\
					& $N_{\rm eff}$                            & $\mathcal{U}$[0\,,\,6] \\                   
\hline\hline
\end{tabular}
\label{table2}
\end{center}	
\end{table}

\begin{table*}[htbp]
\renewcommand\arraystretch{1.5}
\centering
\caption{The 1$\sigma$ confidence regions (or 2$\sigma$ upper limits) of cosmological parameters obtained by the DESI, CMB, DESY5, PantheonPlus, and DESY1 data for the $\Lambda$CDM+$\sum m_\nu+N_\mathrm{eff}$, $w$CDM+$\sum m_\nu+N_\mathrm{eff}$, and $w_0w_a$CDM+$\sum m_\nu+N_\mathrm{eff}$ models. For the parameter $\sum m_\nu$, central values cannot be determined in most cases, and we provide the $2\sigma$ upper limits. Here, $H_{0}$ is in units of ${\rm km}~{\rm s}^{-1}~{\rm Mpc}^{-1}$.}
\resizebox{\textwidth}{!}{%
\setlength{\tabcolsep}{4pt}
\label{table: St}
\begin{tabular}{lccccccc}
\hline\hline
Model/Dataset           & $H_0$ & $\Omega_{\rm m}$ & $S_8$ & $w\ {\rm or}\ w_0$ & $w_a$ & $\sum m_{\nu}\ [\rm eV]$ & $N_{\rm eff}$ \\
\hline
\multicolumn{8}{l}{$\boldsymbol{\Lambda{\rm CDM}+\sum m_\nu+N_\mathrm{eff}}$} \\
CMB+DESI+DESY5          & $68.37\pm0.90$     & $0.3031 \pm 0.0041$ & $0.8215\pm0.0083$ & --- & --- & $< 0.084$ & $3.10\pm0.16$  \\
CMB+DESI+DESY5+DESY1    & $68.13\pm0.90$     & $0.3016 \pm 0.0040$ & $0.8135\pm0.0078$ & --- & --- & $< 0.085$ & $3.02\pm0.16$  \\
\hline
\multicolumn{8}{l}{$\boldsymbol{w{\rm CDM}+\sum m_\nu+N_\mathrm{eff}}$} \\
CMB+DESI+DESY5          & $68.10\pm1.00$ & $0.3084\pm0.0051$ & $0.8226\pm0.0086$ & $-0.956\pm0.023$ & --- & $<0.074$ & $3.24^{+0.18}_{-0.21}$ \\
CMB+DESI+DESY5+DESY1    & $67.78\pm0.95$ & $0.3074\pm0.0049$ & $0.8152\pm0.0077$ & $-0.955\pm0.022$ & --- & $<0.065$ & $3.14\pm0.18$ \\
\hline
\multicolumn{8}{l}{$\boldsymbol{w_0w_a{\rm CDM}+\sum m_\nu+N_\mathrm{eff}}$} \\
CMB+DESI+DESY5                  & $66.66^{+0.71}_{-0.92}$ & $0.3187\pm 0.0058$ & $0.8288\pm 0.0086$ & $-0.753^{+0.054}_{-0.063}$ & $-0.88^{+0.27}_{-0.22}$ & $<0.141$ & $2.85^{+0.40}_{-0.17}$ \\
CMB+DESI+DESY5+DESY1            & $66.54\pm 0.56$ & $0.3187\pm 0.0058$ & $0.8199\pm 0.0081$ & $-0.745^{+0.056}_{-0.063}$ & $-0.91^{+0.26}_{-0.22}$ & $0.098^{+0.016}_{-0.037}$ & $2.46^{+0.60}_{-0.24}$ \\
CMB+DESI+PantheonPlus+DESY1     & $67.37^{+0.77}_{-0.86}$ & $0.3099\pm 0.0057$ & $0.8182\pm 0.0084$ & $-0.845\pm 0.058$ & $-0.58^{+0.24}_{-0.21}$ & $<0.144$ & $2.89^{+0.28}_{-0.18}$ \\
\hline
\end{tabular}
}
\end{table*}

In cosmology, the total energy density of massive neutrinos is given by \cite{WMAP:2010qai}
\begin{equation}
\rho_\nu(a) = \frac{a^{-4}}{\pi^2} \int \frac{q^2\mathrm{d}q}{\mathrm{e}^{q/T_{\nu0}}+1}\sum_i \sqrt{q^2+m_{i}^2a^2},
\label{eq1}
\end{equation}
where $q$ is the comoving momentum, $T_{\nu0}=(4/11)^{1/3}T_\mathrm{cmb}=1.945\,\mathrm{K}$ and $T_\mathrm{cmb}=2.725\,\mathrm{K}$ represent the present neutrino and CMB temperature, respectively. $m_i\,(i=1,2,3)$ are the mass of each neutrino species.

In the early universe, neutrinos were relativistic, thereby $\rho_{\nu}$ is proportional to the photon energy density ($\rho_{\gamma}$), given by \cite{WMAP:2010qai}
\begin{equation}
\rho_{\nu}(a) \to N_{\rm eff} \frac{7}{8} \left(\frac{4}{11}\right)^{4/3} \rho_\gamma(a),
\label{eq2}
\end{equation}
where $N_\mathrm{eff}$ is the effective number of relativistic neutrino species and in the Standard Model, $N_\mathrm{eff}=3.044$ \cite{Akita:2020szl,Froustey:2020mcq,Bennett:2020zkv}. Therefore, Eq.~(\ref{eq1}) can be rewritten as
\begin{equation}
\rho_\nu(a) = N_{\rm eff} \frac{7}{8} \left(\frac{4}{11}\right)^{4/3} \rho_\gamma(a) \sum_i g\left(\frac{m_i}{T_{\nu0}}a\right),
\label{eq3}
\end{equation}
where 
\begin{equation}
g(y) \equiv \frac{40}{7\pi^4}\int_0^\infty \frac{x^2\sqrt{x^2+y^2}}{\mathrm{e}^x+1} \mathrm{d}x.
\label{eq4}
\end{equation}
It can be demonstrated that when $a\to\infty$ (i.e., late-time universe), Eq.~(\ref{eq3}) tends asymptotically to
\begin{equation}
\rho_{\nu}(a) \to \frac{\sum m_{\nu}}{93.14h^2~\rm eV} \rho_\mathrm{crit,0}a^{-3},
\end{equation}
where $h=H_0/(100~\mathrm{km~s^{-1}~Mpc^{-1}})$ is the dimensionless Hubble constant. $\rho_\mathrm{crit,0}=3H_0^2/8\pi G$ is the current critical density.

In a flat Friedmann-Robertson-Walker universe, the dimensionless Hubble parameter can be written as
\begin{align}
E^2(a)&\equiv\frac{H^2(a)}{H_0^2}=(\Omega_\mathrm{b}+\Omega_\mathrm{c})a^{-3}+\Omega_\mathrm{de}f(a)\nonumber\\
&+\Omega_\gamma a^{-4}\biggl[1+N_{\rm eff} \frac{7}{8} \left(\frac{4}{11}\right)^{4/3} \sum_i g\left(\frac{m_i}{T_{\nu0}}a\right)\biggr],
\end{align}
where $\Omega_{\gamma}$, $\Omega_\mathrm{b}$, $\Omega_\mathrm{c}$, and $\Omega_{\rm de}$ represent the current density parameters of photon, baryon, CDM, and DE, respectively, and $\Omega_\gamma = 2.469\times10^{-5}/h^2$ for $T_\mathrm{cmb}=2.725\,\mathrm{K}$. Here, $f(a)$ is the normalized $a$-dependent density of DE, given by
\begin{equation}
f(a) = \exp\left( -3 \int_{0}^{\ln a} [1 + w(a')] {\rm d}\ln a' \right),
\end{equation}
where $w(a)$ is the EoS of DE. In this work, we consider three cosmological models, including the $\Lambda {\rm CDM}+\sum m_\nu+N_\mathrm{eff}$ model with $w=-1$, $w {\rm CDM}+\sum m_\nu+N_\mathrm{eff}$ model with a constant $w$, and $w_0w_a {\rm CDM}+\sum m_\nu+N_\mathrm{eff}$ model, which adopts the Chevallier-Polarski-Linder (CPL) parameterization form, $w(a) = w_0 + w_a(1-a)$~\cite{Chevallier:2000qy,Linder:2002et}.

Table~\ref{table2} summarizes the full set of free parameters and the uniform priors adopted in the analysis. We compute the theoretical models using the {\tt CAMB} code \cite{Lewis:1999bs,Howlett:2012mh} and use the publicly available sampler {\tt Cobaya}~\cite{Torrado:2020dgo} to perform Markov Chain Monte Carlo (MCMC) \cite{Lewis:2002ah,Lewis:2013hha} analysis. We assess the convergence of the MCMC chains using the Gelman-Rubin statistics quantity $R - 1 < 0.02$ \cite{Gelman:1992zz} and the MCMC chains are analyzed using the public package {\tt GetDist}~\cite{Lewis:2019xzd}.

In our main analysis, we consider a joint analysis of the following datasets: (i) \textbf{\texttt{CMB}:} The CMB likelihoods include the temperature and polarization measurements (TT, TE, and EE spectra) from Planck 2018~\cite{Efstathiou:2019mdh,Planck:2018vyg,Planck:2019nip,Rosenberg:2022sdy}, as well as the NPIPE PR4 Planck CMB lensing reconstruction~\cite{Carron:2022eyg} and Data Release 6 of the Atacama Cosmology Telescope~\cite{ACT:2023dou}. (ii) \textbf{\texttt{DESI}:} The BAO measurements from DESI DR2 are summarized in Table IV of Ref.~\cite{DESI:2025zgx}. (iii) \textbf{\texttt{PantheonPlus}:} The PantheonPlus comprises 1550 spectroscopically confirmed supernovae (SNe) from 18 different surveys, covering $0.01 < z < 2.26$~\cite{Brout:2022vxf}. (iv) \textbf{\texttt{DESY5}:} The DESY5 sample comprises 1829 photometrically classified SNe with redshifts in the range $0.025 < z < 1.3$. (v) \textbf{\texttt{DESY1}:} The DESY1 weak lensing data are based on the analysis of 26 million source galaxies and 6.5 million lens galaxies over a 1321 $\mathrm{deg}^2$ footprint~\cite{DES:2017myr,DES:2018zzu}.

\section{Results and discussions}

\begin{figure}[htbp]
\centering
\includegraphics[scale=0.45]{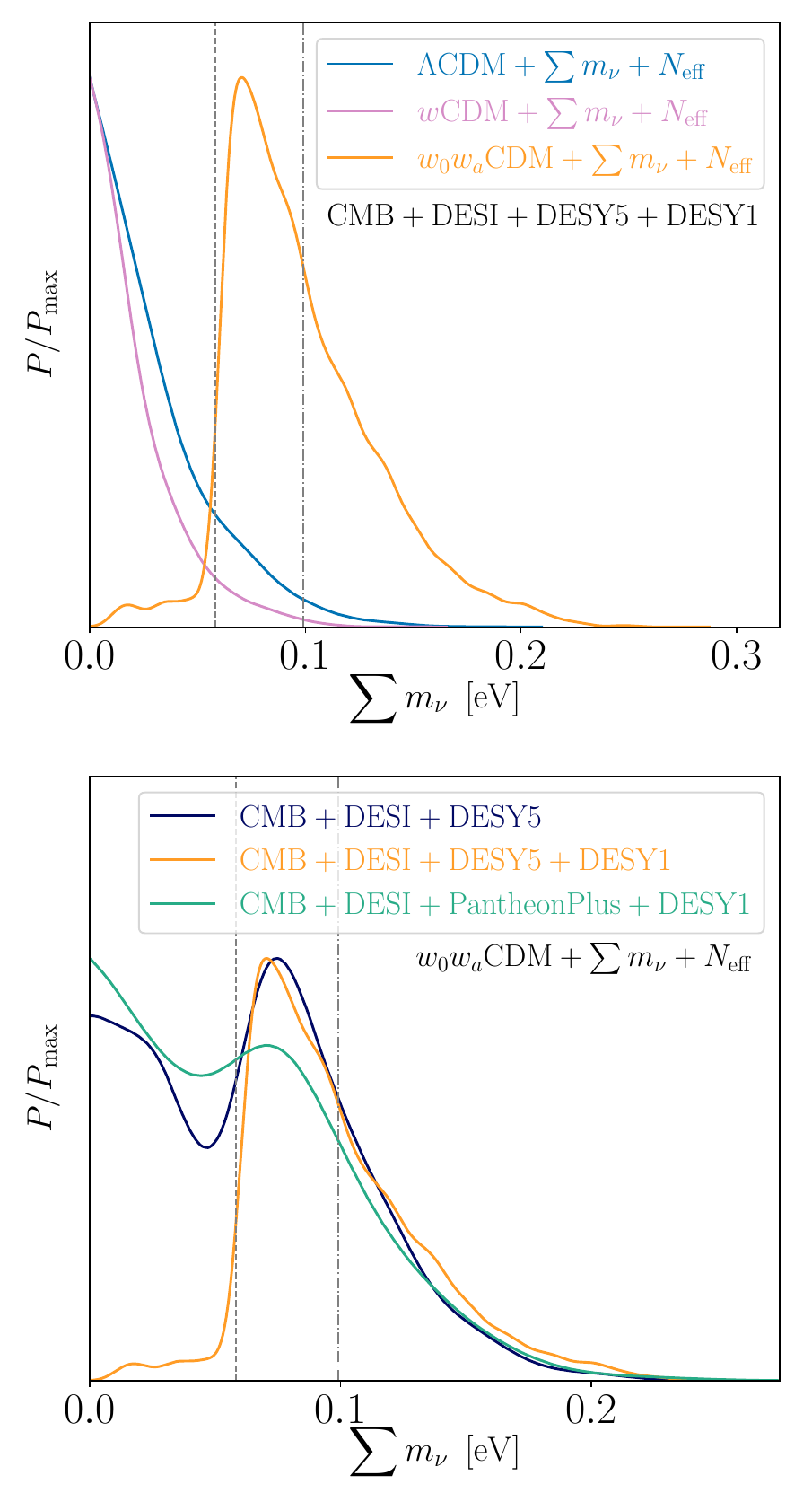}
\caption{\label{fig1} The marginalized 1D posterior distributions on $\sum m_{\nu}$ using DESI, CMB, DESY5, PantheonPlus, and DESY1 data. The dashed and dash-dotted lines represent the lower bounds of neutrino mass in the NH and IH, respectively. \emph{Upper panel}: A comparison of the 1D marginalized posterior distributions for $\sum m_{\nu}$ in the $\Lambda$CDM+$\sum m_\nu+N_\mathrm{eff}$, $w$CDM+$\sum m_\nu+N_\mathrm{eff}$, and $w_0w_a$CDM+$\sum m_\nu+N_\mathrm{eff}$ models using CMB+DESI+DESY5+DESY1 data. \emph{Lower panel}: A comparison of the 1D marginalized posterior distributions for $\sum m_{\nu}$ in the $w_0w_a$CDM+$\sum m_\nu+N_\mathrm{eff}$ models using CMB+DESI+DESY5, CMB+DESI+DESY5+DESY1, and CMB+DESI+PantheonPlus+DESY1 data.}
\end{figure}

\begin{figure*}[htbp]
\includegraphics[scale=0.45]{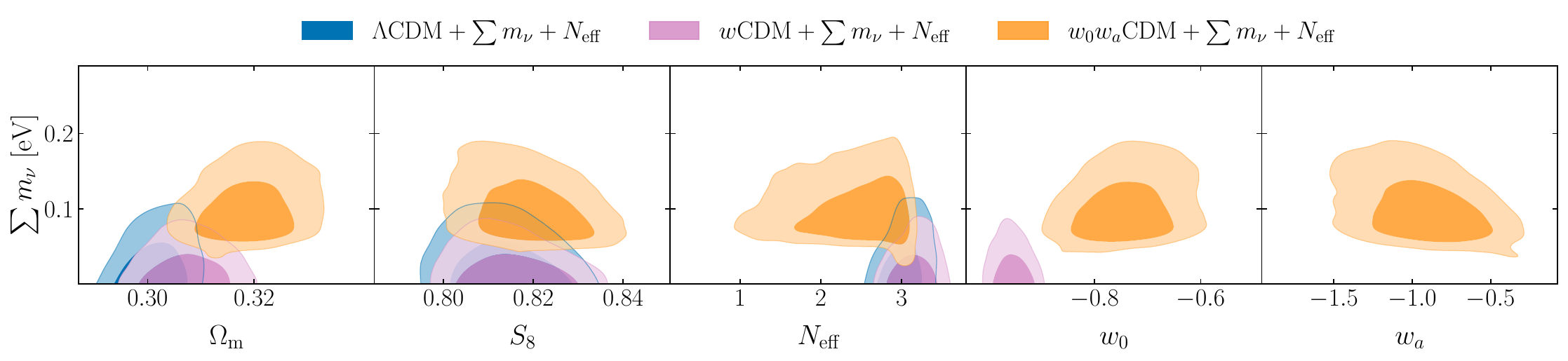}
\centering
\caption{\label{fig2} A comparison of the two-dimensional marginalized contours between $\sum m_{\nu}$ and various other cosmological parameters using CMB+DESI+DESY5+DESY1 data.}
\end{figure*}

\begin{figure}[htbp]
\includegraphics[scale=0.43]{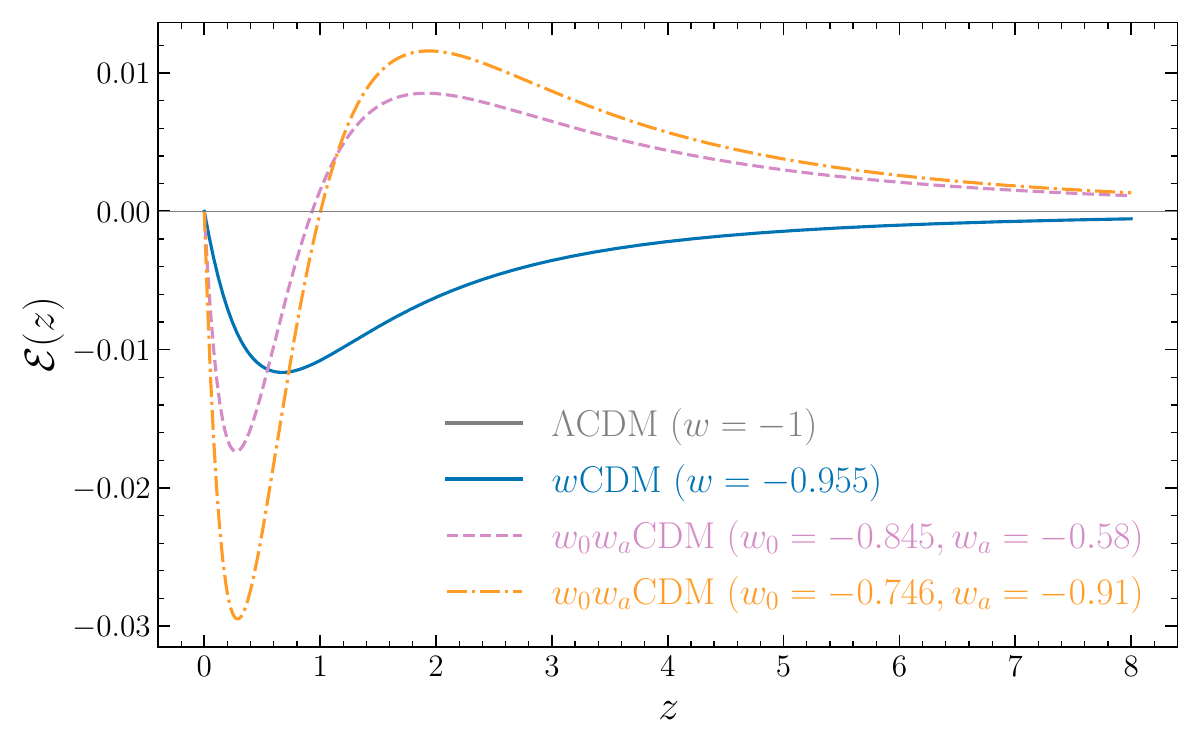}
\centering
\caption{\label{fig3} The redshift evolution of $\mathcal{E}(z)$. The gray solid curve denotes the $\Lambda$CDM model, while the blue solid curve corresponds to the $w$CDM model constrained by CMB+DESI+DESY5+DESY1 data ($w=-0.955$). The pink dashed curve represents the $w_0w_a$CDM model using CMB+DESI+PantheonPlus+DESY1 ($w_0=-0.845$, $w_a=-0.58$). The orange dash-dotted curve is the $w_0w_a$CDM constraints from CMB+DESI+DESY5+DESY1 ($w_0=-0.746$, $w_a=-0.91$).}
\end{figure}

In this section, we present the main parameter constraint results in Table~\ref{table: St} and Figs.~\ref{fig1} and \ref{fig2}. Figure~\ref{fig3} shows the relative variation of the dimensionless Hubble expansion rate.

For the $\Lambda\mathrm{CDM}+\sum m_\nu+N_{\rm eff}$ model, we obtain an upper limit on the total neutrino mass of $\sum m_\nu<0.084\ \mathrm{eV}$ using CMB+DESI+DESY5 data, and $\sum m_\nu<0.085\ \mathrm{eV}$ when DESY1 is included. The inclusion of weak lensing data slightly raises the upper limit on $\sum m_\nu$, since the lower clustering amplitude ($S_8$) inferred from weak lensing data allows for larger neutrino masses. Building on this, when a constant DE EoS is introduced, the $w\mathrm{CDM}+\sum m_\nu+N_{\rm eff}$ model yields $\sum m_\nu<0.065\ \mathrm{eV}$ and $w=-0.955\pm0.022$ using CMB+DESI+DESY5+DESY1 data. This indicates that a constant EoS of DE with $w>-1$ lowers the neutrino mass upper limit, as illustrated in the upper panel of Fig.~\ref{fig1}, which is in agreement with previous studies~\cite{Zhang:2015uhk,Vagnozzi:2018jhn,Du:2024pai}.

When we further consider a dynamically evolving DE, the CMB+DESI+DESY5+DESY1 data yield $\sum m_\nu = 0.098^{+0.016}_{-0.037}\,\mathrm{eV}$, indicating a $2.7\sigma$ measurement of a neutrino mass (for further robustness checks regarding this result, see Appendix~\ref{appendixA}). This high-confidence measurement primarily arises for the following reasons:
\begin{itemize}
\item DESI favors a dynamical DE EoS that evolves from $w< -1$ at early times to $w> -1$ at late times. This evolving behavior, which leads to a larger neutrino mass, is the primary reason for the high-confidence measurement reported in our analysis, as concluded in our previous systematic study~\cite{Zhang:2015uhk,Zhao:2016ecj,Zhang:2017rbg,Zhang:2020mox,Du:2024pai}. Meanwhile, the detection of a non-zero neutrino mass is strongly influenced by evidence for the dynamical evolution of DE. As shown in Fig.~\ref{fig2}, $\sum m_\nu$ is anti-correlated with $w_a$ and positively correlated with $w_0$. The CMB+DESI+DESY5+DESY1 data give $w_0=-0.745^{+0.056}_{-0.063}$ and $w_a = -0.91^{+0.26}_{-0.22}$, corresponding to evidence for a dynamically evolving DE at the $\sim 4\sigma$ level, thereby favoring a larger neutrino mass. In contrast, replacing DESY5 with PantheonPlus yields CMB+DESI+PantheonPlus+DESY1 results of $\sum m_\nu < 0.144\,\mathrm{eV}$, where the evidence for DE dynamics drops to $\sim2\sigma$ ($w_0 = -0.845 \pm 0.058$, $w_a = -0.58^{+0.24}_{-0.21}$), leading only to an upper limit on the neutrino mass. We further test this conclusion by introducing priors with varying evidences for dynamically evolving DE; see Appendix~\ref{appendixB} for a detailed discussion.

\item Treating $N_\mathrm{eff}$ as a free parameter also impacts the neutrino mass measurement. In the $w_0w_a\mathrm{CDM}+\sum m_\nu+N_{\rm eff}$ model using CMB+DESI+DESY5+DESY1 data, we obtain $N_{\mathrm{eff}} = 2.46^{+0.60}_{-0.24}$. Although this value is consistent with the particle-physics standard value $3.044$ at the $1\sigma$ level, its central value lies below the standard value. One possible physical explanation for a reduced $N_{\mathrm{eff}}$ is a low reheating temperature in the early universe\footnote{The CMB and BBN observations place strong constraints on the lower limit of the reheating temperature, typically requiring $T_\mathrm{RH} \gtrsim  4-5\,\mathrm{MeV}$~\cite{deSalas:2015glj,Hasegawa:2019jsa}. Nevertheless, in such scenarios, there still exists an allowed parameter space where $N_\mathrm{eff}$ can be significantly smaller than 3.044~\cite{deSalas:2015glj}.}, which prevents active neutrinos from fully thermalizing to the standard Fermi-Dirac distribution~\cite{Kawasaki:2000en,Giudice:2000ex}. Some studies have shown that in such scenarios the late-time constraint on $\sum m_\nu$ can be substantially relaxed~\cite{deSalas:2015glj}. This follows because a reduced neutrino number density, or a colder neutrino population (leading to $N_{\rm eff}<3.044$), requires a larger total neutrino mass to produce the same late-time effects on background expansion and structure formation.

\item {The inclusion of weak lensing data also allows for a larger neutrino mass. Specifically, $\sum m_\nu$ is anti-correlated with $S_8$, and the lower $S_8$ values preferred by the DESY1 weak lensing data also push the neutrino mass higher. Without weak lensing, CMB+DESI+DESY5 only impose an upper limit of $\sum m_\nu < 0.141\,\mathrm{eV}$, with a multi-peaked one-dimensional (1D) marginalized posterior distribution, as shown in lower panel of Fig.~\ref{fig1}. The inclusion of DESY1 suppresses the lower-mass peak of the posterior, thereby providing evidence for a non-zero neutrino mass. }
\end{itemize}

Notably, the DESI collaboration recently found that, under a hypothetical unphysical negative neutrino-mass scenario (i.e., one whose effects on the background expansion and structure formation are opposite to those of a positive-mass neutrino), nearly all cosmological observations preferentially indicate a negative neutrino mass, producing a tension of up to $3\sigma$ with neutrino-oscillation results. Dynamical DE can effectively relieve this tension~\cite{Elbers:2025vlz}. The DESI analysis carried out a systematic investigation of neutrino cosmology using a variety of data combinations (DESI DR1 full-shape, DESI DR2 BAO, CMB, and several SN compilations), a range of DE models, and different neutrino-physics assumptions. In contrast, our analysis concentrates on the case of dynamical DE while treating both $\sum m_{\nu}$ and $N_{\mathrm{eff}}$ as free parameters, and additionally includes weak gravitational lensing data to further broaden the neutrino mass, enabling a cosmological measurement of a non-zero, positive neutrino mass. As discussed above, we obtain a detection of non-zero neutrino mass at up to $2.7\sigma$, which is broadly consistent with neutrino-oscillation experiments and demonstrates the considerable potential of cosmological observations to measure the neutrino mass. While the current analysis is phenomenologically driven, these robust observational preferences underscore the imperative for future theoretical works to construct explicit beyond-$\Lambda$CDM models. Investigating scenarios such as those involving interactions between dark matter and neutrinos~\cite{Brax:2023tvn,Zu:2025lrk}, will be crucial to elucidate the underlying physical mechanisms driving these data anomalies.

Next, we further illustrate the impact of DE on neutrino mass measurements by examining the evolution of the late-time relative expansion rate. The CMB TT power spectrum data tightly constrain the position and amplitude of the first acoustic peak. The determination of the first peak depends precisely on the angular size of the sound horizon at decoupling, $\Theta_{\rm s}\equiv r_{\rm s}/D_\mathrm{A}(z_\mathrm{LS})$, and the redshift of matter-radiation equality, $z_{\rm eq}$. Here, the sound horizon at decoupling $r_{\rm s}$ is essentially fixed by the physics of pre-recombination periods, while DE variations are only relevant at late times; therefore, $r_{\rm s}$ is unaffected by DE \cite{Vagnozzi:2018jhn}. To keep both the angular diameter distance of last scattering $D_\mathrm{A}(z_\mathrm{LS})$ and $z_{\rm eq}$ approximately fixed, any change in DE must be compensated by shifts in other cosmological parameters. $D_\mathrm{A}(z_\mathrm{LS})$ is given by
\begin{equation}\label{eq8}
D_\mathrm{A}(z_\mathrm{LS}) = \frac{c}{H_0(1+z_\mathrm{LS})} \int_{0}^{z_\mathrm{LS}} \frac{1}{E(z)} \,\mathrm{d}z,
\end{equation}
where $z_{\rm LS}$ is the redshift of last scattering. We then define
\begin{equation}
\mathcal{E}(z) =
\left.
\frac{E(z)_{\Lambda\mathrm{CDM}}}{E(z)_{w\mathrm{CDM}/ w_0w_a\mathrm{CDM}}}
\right|_{\Omega_{\rm b},\,\Omega_{\rm c},\,\Omega_{\rm de},\,\sum m_\nu,\,N_{\rm eff}}
\;-\;1,
\end{equation}
where the notation $\left.\right|_{\Omega_{\rm b},\,\Omega_{\rm c},\,\Omega_{\rm de},\,\sum m_\nu,\,N_{\rm eff}}$ indicates that these parameters are held fixed to the best-fit values obtained in the $\Lambda\mathrm{CDM}+\sum m_\nu+N_{\mathrm{eff}}$ model from the $\mathrm{CMB}+\mathrm{DESI}+\mathrm{DESY5}+\mathrm{DESY1}$ data, with $\Omega_{\rm b} = 0.048$, $\Omega_{\rm c} = 0.253$, $\Omega_{\rm de} = 0.699$, $\sum m_\nu = 0.085$ eV (the $2\sigma$ upper limit), and $N_\mathrm{eff} = 3.02$.

The redshift evolution of $\mathcal{E}(z)$ is shown in Fig.~\ref{fig3}. It is clear that when the curve lies above zero, DE variations increase the integral in Eq.~(\ref{eq8}), and when it lies below zero, they decrease it. In the $w$CDM model, $w>-1$ yields a relative expansion rate higher than that in $\Lambda$CDM, resulting in $\mathcal{E}(z)<0$ (blue solid line). To keep $D_{\rm A}(z_{\rm LS})$ unchanged, directly reducing $\Omega_{\rm b}$, $\Omega_{\rm c}$, or $N_{\rm eff}$ is undesirable because each shifts $z_{\rm eq}$ and would introduce unwanted changes elsewhere in the CMB power spectrum. Instead, lowering $\sum m_\nu$, whose effect on $z_{\rm eq}$ is minimal, is the most economical option for compensating the reduced integral. Consequently, introducing a constant EoS of DE with $w>-1$ tightens the upper limit on $\sum m_\nu$.

For the $w_0w_a$CDM model, current data strongly prefer a crossing DE EoS that evolves from $w<-1$ at early times to $w>-1$ at late times. As shown in Fig.~\ref{fig3}, this DE dynamics leads to a significant alteration in the late-time relative expansion rate of the universe: it exceeds that of the $\Lambda$CDM model for $0<z\lesssim1$, whereas it falls below it for $z\gtrsim1$. It should be emphasized that CMB observations constrain the position of the first acoustic peak with extremely high precision, which requires the angular diameter distance to the last scattering surface, $D_{\rm A}(z_{\rm LS})$, to remain highly stable. Since the integration range for $1/E(z)$ in Eq.~(\ref{eq8}) extends to $z_{\rm LS}\approx1090$, the negative deviation in the $z\gtrsim1$ region dominates the entire geometric integral, leading to a substantial overall increase in the integral value compared to the standard model.

To maintain $D_{\rm A}(z_{\rm LS})$ constant to fit the CMB data, the model must introduce a corresponding physical compensation effect. If one directly adjusts $\Omega_{\rm m}$ or $H_0$ to compensate for the integral change, it would inevitably alter the redshift of matter-radiation equality, $z_{\rm eq}$, thereby spoiling the spectral structure of the CMB power spectrum. Therefore, increasing the total neutrino mass $\sum m_\nu$, which has a minimal impact on $z_{\rm eq}$, becomes the most natural choice to offset this geometric effect of DE dynamics. This is highly consistent with the parameter degeneracy mechanisms revealed in our previous series of studies \cite{Zhang:2015uhk,Zhao:2016ecj,Zhang:2017rbg,Zhang:2020mox,Du:2024pai}.

Moreover, such crossing DE dynamics (the joint contribution of $w_0$ and $w_a$) tends to yield a lower $H_0$ in the inference, which not only suppresses $N_{\rm eff}$ but also slightly alters the theoretical expectation of $z_{\rm eq}$. Thus, the dynamically evolving DE in the $w_0w_a$CDM model intrinsically drives an increase in $\sum m_\nu$ and a decrease in $N_{\rm eff}$, a physical picture that corroborates the latest conclusions of the DESI collaboration \cite{DESI:2024mwx,DESI:2025zgx,Elbers:2025vlz}. Finally, as the evidence supporting dynamically evolving DE strengthens (i.e., the transition from the pink dashed line to the orange dash-dot line in Fig.~\ref{fig3}, reflecting a larger deviation of $w_0, w_a$ from $-1, 0$), the evolutionary deviations in the relative expansion rate become more drastic. This demands a larger $\sum m_\nu$ and a smaller $N_{\rm eff}$ to absorb the late-time geometric effects brought about by the variation in the DE EoS, thereby providing high-confidence observational evidence for a non-zero neutrino mass at the physical level.

Finally, it is worth noting that Ref.~\cite{RoyChoudhury:2025dhe} also reported a $2\sigma$ measurement of $\sum m_\nu = 0.190 \pm 0.088\,\mathrm{eV}$ in an extended cosmological model with 12 parameters, including dynamical DE, the running of the scalar spectral index ($\alpha_s$), and the scaling of the lensing amplitude ($A_{\rm lens}$). Compared with this study, we only consider a dynamical DE scenario including two important neutrino parameters ($\sum m_\nu$ and $N_{\rm eff}$), motivated by the key conclusions of our earlier series of works \cite{Zhang:2015uhk,Zhao:2016ecj,Zhang:2017rbg,Zhang:2020mox,Du:2024pai} and DESI's preference for dynamical DE with its EoS evolving from $w<-1$ to $w>-1$. As anticipated, relative to analyses incorporating other extensions, we achieve a higher statistic significance for the neutrino mass measurement. Furthermore, our result lies below but close to the IH bound and therefore does not support the IH, which is consistent with the conclusions of particle physics experiments \cite{DeSalas:2018rby,Gariazzo:2018pei}.

\section{Conclusion}

In this paper, we investigate the cosmological implications of simultaneously including $\sum m_\nu$ and $N_{\rm eff}$ within the framework of dynamical DE. We further clarify the specific impact of dynamical DE on neutrino mass measurements from the perspectives of parameter degeneracies and the relative expansion rate.

Our joint analysis of DESI DR2, CMB, DESY5, and DESY1 data yields $\sum m_\nu = 0.098^{+0.016}_{-0.037}\,\mathrm{eV}$, indicating a measurement of a non-zero neutrino mass at the $2.7\sigma$ level within the $w_0w_a\mathrm{CDM}+\sum m_\nu+N_{\rm eff}$ model. This high-confidence measurement primarily arises from these factors: (i) The current DESI results favor a dynamical DE EoS evolving from $w<-1$ to $w>-1$, which tends to increase the neutrino mass. When DESY5 is replaced with PantheonPlus, the evidence for a positive neutrino mass disappears, as the evidence for dynamical DE becomes weaker. Furthermore, only upper limits are obtained in the $\Lambda$CDM and $w$CDM frameworks. The overall outcomes highlight the conclusion from our previous systematic studies~\cite{Zhang:2015uhk,Zhang:2017rbg}. (ii) Allowing $N_{\mathrm{eff}}$ to vary (with a best-fit value below the standard value 3.044) also requires a larger $\sum m_\nu$ to compensate. (iii) The inclusion of weak lensing data, which prefers a relatively lower $S_8$, also suppresses the possibility of a lower neutrino mass. Finally, we test the robustness of our results by investigating how DE affects the detection of neutrino mass and alters the evidence for dynamical DE, through the evolution of the late-time relative expansion rate. We find that as the evidence for dynamical DE strengthens, the deviations in the relative expansion rate become more pronounced, requiring a larger $\sum m_\nu$ to compensate for the DE changes. This serves as the most economical way to offset the integral variation and consequently enhances the evidence for a non-zero neutrino mass.

Our results indicate that this represents a highly compelling cosmological scenario, and that the DESI dataset has inaugurated a new era for measuring neutrino mass through cosmological observations. With the forthcoming, more complete DESI data releases and the inclusion of full-shape modeling of the power spectrum, stronger preferences for dynamical DE are likely to emerge, thereby offering promising prospects for further revealing cosmological evidence of a non-zero neutrino mass.

\section*{Acknowledgments}
We thank Sheng-Han Zhou, Jia-Le Ling, and Yi-Min Zhang for their helpful discussions. This work was supported by the National Natural Science Foundation of China (Grants Nos. 12533001, 12575049, and 12473001), the National SKA Program of China (Grants Nos. 2022SKA0110200 and 2022SKA0110203), the China Manned Space Program (Grant No. CMS-CSST-2025-A02), and the National 111 Project (Grant No. B16009).

\section*{Conflict of Interest}
The authors declare that they have no conflict of interest.

\bibliography{main}

\newpage
\appendix

\section{Robustness check of the positive neutrino mass}\label{appendixA}

\begin{figure}[htbp]
\includegraphics[scale=0.85]{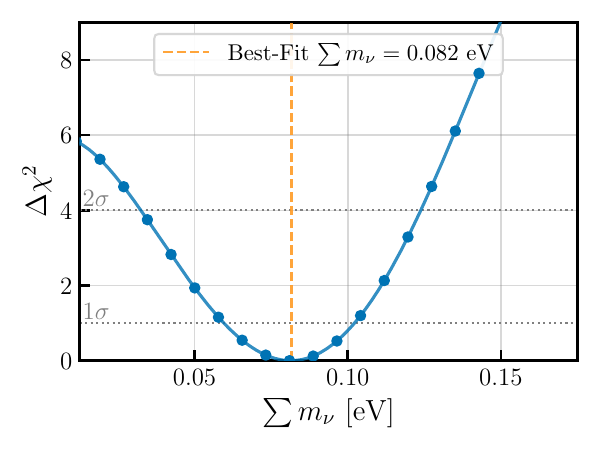}
\centering
\caption{\label{fig4} Using the CMB+DESI+DESY5+DESY1 data, the profile likelihood for the parameter $\sum m_\nu$ for the $w_0w_a{\rm CDM}+\sum m_\nu+N_\mathrm{eff}$ model.}
\end{figure}

\begin{figure}[htbp]
\includegraphics[scale=0.50]{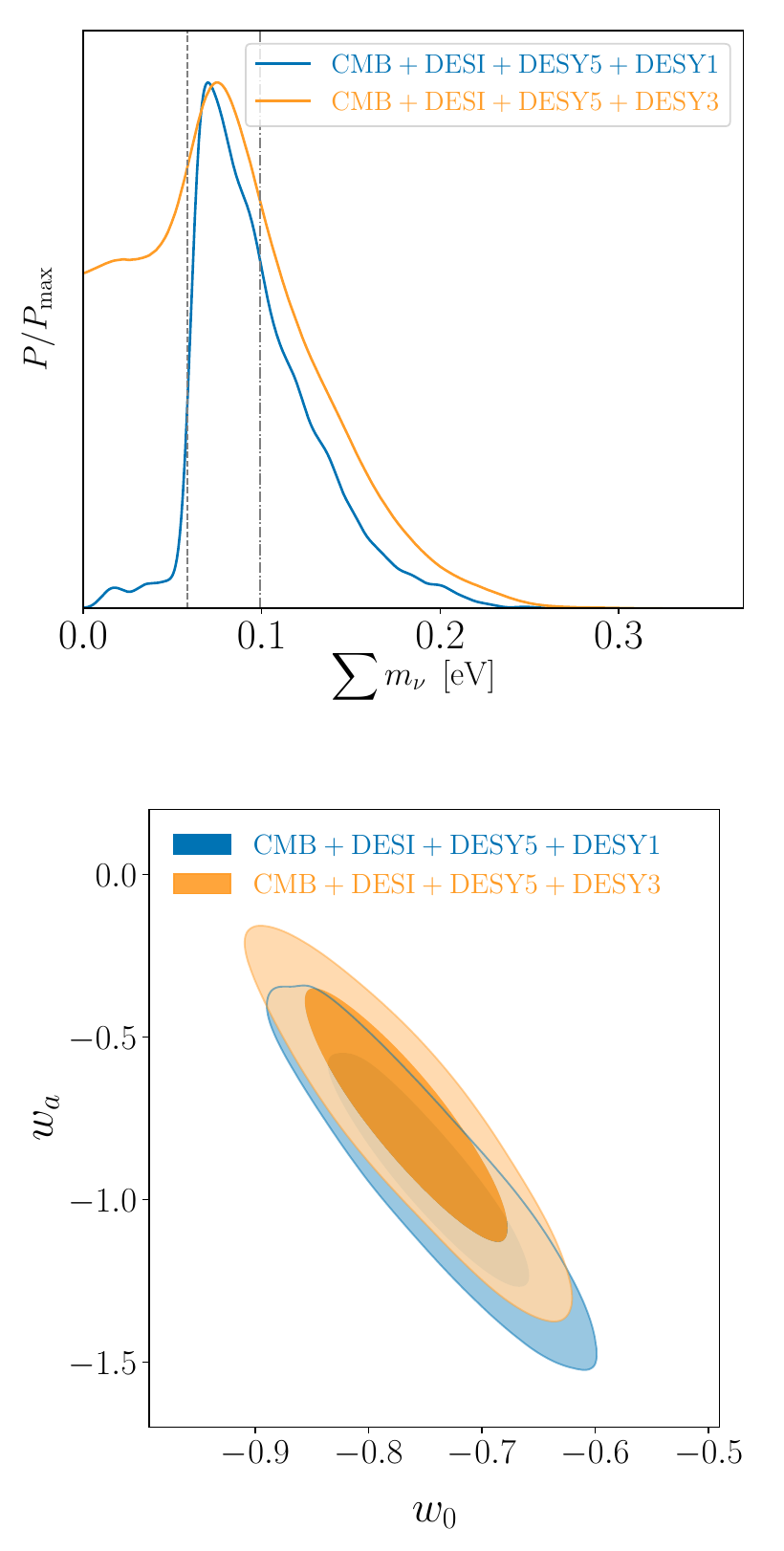}
\centering
\caption{The 1D posterior distribution of $\sum m_\nu$ (upper) and the 2D marginalized contours in the $w_0-w_a$ plane (lower) using the CMB+DESI+DESY5+DESY1 and CMB+DESI+DESY5+DESY3 data, respectively.}
\label{fig5}
\end{figure}

In this Appendix, we comprehensively test the robustness of our preference for a positive neutrino mass by examining the stability of the results against prior volume effects, boundary truncations, and dataset variations. In parameter inference, traditional MCMC methods can sometimes be subject to Bayesian prior volume effects, which may cause the peak of the posterior distribution to deviate from the true best-fit point. To rigorously examine the statistical robustness of our $2.7\sigma$ preference for a positive neutrino mass, we employ the profile likelihood (PL) method, which is independent of priors. We fix $\sum m_\nu$ at a series of discrete values within the range of $[0, 0.15]~\mathrm{eV}$, and at each fixed point, we compute the global maximum likelihood for the remaining cosmological parameters. As illustrated in Fig.~\ref{fig4}, the minimum $\chi^2$ occurs at $\sum m_\nu = 0.082~\mathrm{eV}$, which is remarkably close to the central value of $0.098~\mathrm{eV}$ given by the MCMC analysis. More importantly, the $\Delta \chi^2$ curve rises significantly towards the low-mass end, reaching $\Delta \chi^2 \approx 6$ as $\sum m_\nu \to 0$. Therefore, the current joint datasets possess a genuine physical preference for a positive neutrino mass, rather than being a statistical artifact.

To examine the robustness of our results with respect to the choice of datasets, we conducted a comparative analysis by replacing DESY1 with DESY3. Note that the DESY3 weak lensing data are based on the analysis of approximately 100 million source galaxies and 10 million lens galaxies over a 4143 $\mathrm{deg}^2$ footprint, providing updated cosmic shear and galaxy clustering observations~\cite{DES:2021wwk,DES:2022ccp}. Using CMB+DESI+DESY5+DESY3, we obtain $\sum m_\nu = 0.081^{+0.028}_{-0.071}~\mathrm{eV}$. As illustrated in the upper panel of Fig.~\ref{fig5}, the peak of the 1D marginalized posterior distribution of the neutrino mass shifts to the left, and the low-mass peak reappears. This is fully consistent and internally coherent with the conclusions presented in our manuscript, namely that the non-zero neutrino mass is primarily driven by the data preference for dynamical DE. Compared to DESY1, although the joint data with DESY3 also yields a relatively lower $S_8 = 0.8179\pm 0.0079$, it provides weaker evidence for the dynamical evolution of DE ($w_0 = -0.765\pm 0.059$, $w_a = -0.75^{+0.27}_{-0.24}$), as shown in the lower panel of Fig.~\ref{fig5}. Based on the relative expansion rate compensation mechanism discussed in detail earlier, the weakened evidence for the dynamical evolution of DE naturally leads to a corresponding decrease in the required $\sum m_\nu$ compensation. This robustness test not only confirms the sensitivity of the results to the datasets but also inversely verifies the degeneracy relationship between dynamical DE and the neutrino mass.

It is worth noting that some recent analyses have allowed for an unphysical negative neutrino mass to examine statistical behaviors~\cite{Craig:2024tky,Green:2024xbb,Elbers:2025vlz}. Under the standard $\Lambda$CDM model, since the peak of the posterior distribution lies in the negative half-axis and deviates from the $\sum m_{\nu}=0$ boundary, such an extension helps reveal potential data inconsistencies or tensions. However, under the $w_{0}w_{a}$ CDM model considered in this work, the peak of the posterior distribution is already on the positive half-axis and deviates from the zero boundary (as demonstrated by our PL and MCMC results, where the peak is located between 0.082 and 0.098 eV). Since the posterior distribution itself is no longer primarily affected by the truncation boundary, extending the parameter space into the unphysical negative region would not yield any substantial changes to the physical inference in the positive domain.

\section{Impact of dynamical DE evidence on neutrino mass}\label{appendixB}

\begin{figure}[htbp]
\includegraphics[scale=0.45]{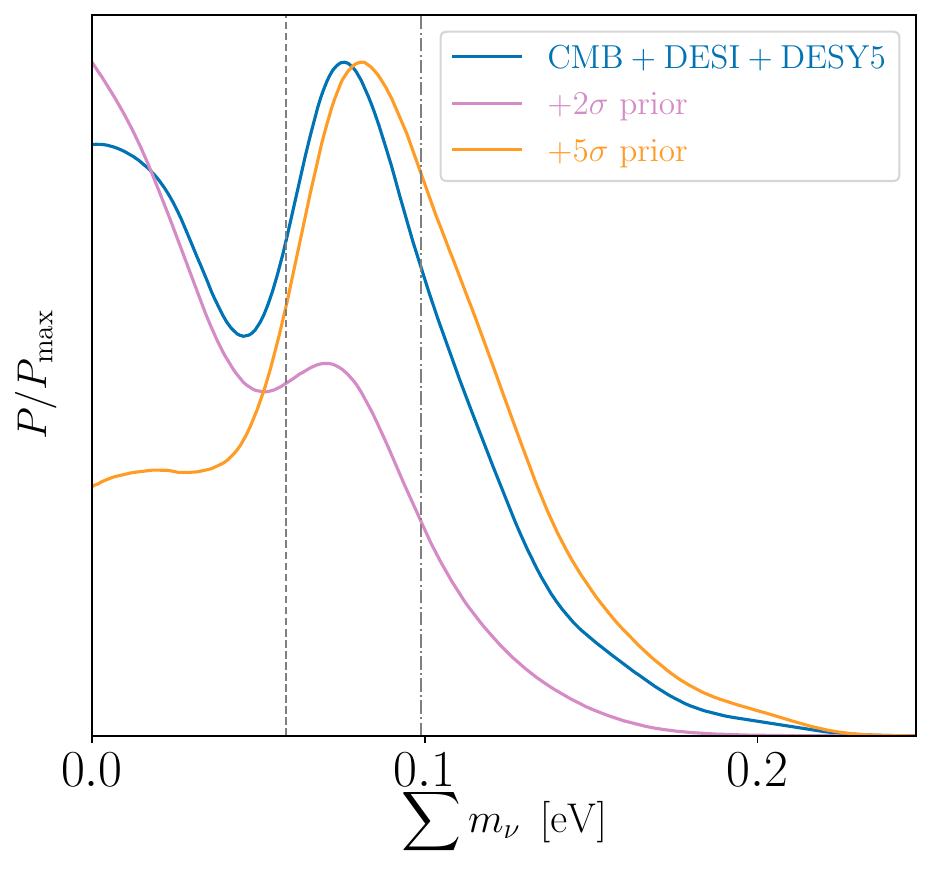}
\centering
\caption{\label{fig6} The 1D marginalized posterior distributions of $\sum m_{\nu}$ in the $w_{0}w_{a}\mathrm{CDM} + \sum m_{\nu} + N_{\rm eff}$ model. The CMB+DESI+DESY5 constraint is regarded as the baseline, shown together with the inclusion of $2\sigma$ and $5\sigma$ priors of dynamical DE.}
\end{figure}

In this Appendix, we report and discuss the impact of the evidence for dynamical DE on neutrino mass constraints. We adopt the constraints on the $w_{0}w_{a}\mathrm{CDM} + \sum m_{\nu} + N_{\rm eff}$ model derived from CMB+DESI+DESY5 data as our baseline. We then impose separately a $2\sigma$ prior ($w_{0}=-0.86\pm0.07$, $w_{a}=-0.50\pm0.25$) and a $5\sigma$ prior ($w_{0}=-0.65\pm0.07$, $w_{a}=-1.25\pm0.25$). The resulting 1D marginalized posterior distributions of $\sum m_{\nu}$ are shown in Fig.~\ref{fig6}. As expected, for the bimodal structure of the CMB+DESI+DESY5 posterior, the $2\sigma$ prior suppresses the high-mass peak, yielding an upper limit of $\sum m_{\nu} < 0.118\ \mathrm{eV}$. Conversely, the $5\sigma$ prior suppresses the low-mass peak, leading to $\sum m_{\nu} = 0.083^{+0.046}_{-0.039}\ \mathrm{eV}$. This further confirms the significant impact of evidence for dynamical DE on neutrino mass measurements and ensures the robustness of our conclusions.

\end{document}